# Structure-Based Super-Resolution Recovery of Three-Photon Quantum States from Two-Fold Coincidences

Dikla Oren[1], Yoav Shechtman[1,2], Maor Mutzafi[1], Yonina C. Eldar[3], Mordechai Segev[1]

[1] *Physics Department and Solid State Institute, Technion, 32000 Haifa, Israel*
[2] *Department of Chemistry, Stanford University, 375 North-South Mall, Stanford, California 94305, USA*
[3] *Electrical Engineering Department, Technion, 32000 Haifa, Israel*

The field of quantum information has been growing fast over the past decade. In particular, optical quantum computation, based on the concepts of KLM[1] and cluster states[2], has witnessed experimental realizations of larger and more complex systems in terms of photon number[3]. Quantum optical systems, which offer long coherence times and easy manipulation of single qubits and photons, allow us to probe quantum properties of the light itself[4] and of the physical systems around it[5]. Recently, a linear scheme for quantum computing, relying on the bosonic nature of particles, has been proposed[6] and realized experimentally with photons[4,7,8]. The ability to efficiently measure superpositions of quantum states consisting of several photons is essential to the characterization of such systems and computation units. In fact, the entire field of quantum information completely relies on the ability to recover quantum states from measurements. However, the characterization of quantum states requires many measurements, and often necessitates complicated measurements schemes; for example, characterizing $m$ qubits requires $2^{2m}$ measurements. Here, we utilize structure, inherent to physically interesting quantum states of light, in order to reduce the complexity in the recovery of a

**quantum state. In particular, we devise a method enabling the recovery of three-photon quantum states (including entangled states) from only two-fold correlation measurements in a single setting. The ability to take two-fold coincidences in a single setup instead of three-fold offers the recovery of the quantum states in far less measurements and in a considerably higher signal-to-noise ratio, because detection of two-photon events is much more likely than that of three-photon ones. The concept suggested here paves the way to further ideas on structure-based super resolution in quantum state tomography, such as recovering a quantum state in an unknown basis in a single setup and recovering the state of several photons without number resolving detectors.**

Quantum information processing and quantum computation have drawn considerable attention both theoretically and experimentally. The concept of a quantum computer, dating back to Feynman's quantum simulator[9], gained much momentum two decades ago, when Shor proposed a quantum algorithm which offers an exponential speedup of prime factoring[10]. Ever since, qubits and quantum gates, the basic building blocks of a universal quantum computer, have been proposed and realized experimentally in a variety of physical systems ranging from trapped ions[11] and photons[12] to nuclear magnetic resonance[13] and cluster states[3]. More recently, non-universal schemes, which still hold purely quantum properties, have been investigated[6,14,15]. One of these schemes, known as Boson Sampling[6], has been realized experimentally in systems with up to 5 photons[4,7,8,16]. Common to all of these quantum systems is the need to recover the quantum state from measurements. Knowing the quantum state is essential to the characterization of prepared states, demonstration of computational units and retrieval of the results of a computation. In order to reveal the complete quantum state, quantum state tomography (QST) is performed. In this process, a measurement corresponding to every single element in the density matrix is repeated many times. In principle, QST yields the full density matrix. However, it suffers from two main drawbacks: (i) the number of required measurements is very large, and (ii) different physical realizations of measurement systems are necessary. More specifically, the density matrix of an $m$ qubit system has $2^{2m}$ complex elements, which, when taking into account the structure of the density matrix, corresponds to $2^{2m}-1$ real parameters required for obtaining a reliable measurement. Each of these parameters is measured by varying the physical system or its coupling to the detectors. This can take, for example, the form of changing the magnetic field locally in every spin position in spin qubits, or inserting and removing optical components such as polarizers and beam splitters in an optical system. When considering photonic quantum systems, coincidences and correlation functions are used to characterize the states. As the number of photons in the system grows, higher order coincidences and correlation functions are needed[4,7,8,16,17]. These issues raise a natural question: Can we characterize a quantum state of $N$ photons from lower order correlations? Furthermore, can we achieve this goal in a single experimental setup, without changing the physical system?

Here, we present a new paradigm to recover quantum states from lower-order coincidence measurements, which are simpler, fewer and are obtained at a higher rate. In particular, we devise a method to recover three-photon quantum states from only two-fold correlation measurements in a single setting. The ability to take two-fold coincidences instead of three-fold offers the recovery of the quantum states in far less measurements and in a considerably higher signal-to-noise ratio (SNR), because two-photon events are much more likely than three-photon ones. For example, in a system with detection quantum efficiency of 10% , for every triplet injected into the system only one in a thousand is detected, whereas 27 pairs are recorded within the same time. This increase in rate improves the SNR of the measurement.

The concept we propose relies on the fact that, in many cases of interest, the quantum information processed in the system has some characteristic structure (i.e., it is not random). This structure stems from the physical nature of the state and from the fact that, often, we are interested in states that are either pure states or pure states that have undergone some degradation but are still close to pure states. In this context, having structure in a signal implies that this signal has a sparse representation in some basis, that is, it can be represented in this basis by only a small number of coefficients. Such a signal is then said to be sparse in that basis[18]. For such cases, where the quantum states are close to pure states, we demonstrate the recovery of three-photon states from only two-fold correlation measurements in a single setup, without the need to change the physical system. We propose a general algorithmic technique to recover the state, provide specific examples with photon-number states and with entangled states, and evaluate the performance of our methodology with respect to sparsity and noise. Finally, we discuss future ideas on how to utilize structure-based concepts even when the sparsity basis is unknown, and related ideas where additional information can be unraveled algorithmically from a partial set of measurements.

We demonstrate the concept of structure-based super-resolution on a specific photonic system: an array of $N_w$ evanescently-coupled optical waveguides (Fig. 1a). This system has been used extensively to study fundamental concepts in both the classical and quantum domains (e.g., Bloch oscillations[19,20], Zener tunneling[21], Shockley states[22], bound states in the continuum[23], Anderson localization[24,25], and topological insulators[26], as mean-field, and in the single-photon regime[5,27–29]). In the context presented here, we use this system due to the following reasons: (1) the field forms an inherently discrete set of modes, which is most suitable for quantum information schemes, (2) the waveguide array is lossless and exhibits no decoherence, (3) the system is simple and experimentally realizable, and (4) the waveguides are coupled to each other, which means that measuring light at the output of any waveguide reveals information about other waveguides as well (Fig. 1b). As will be explained below, the spreading of information among the modes is essential for our approach. A similar technique utilizing a coupled waveguide array in order to recover information at the input of the array has been recently demonstrated in the classical realm[30]. It is important to emphasize that even though we illustrate the idea on this specific system, the concept of structure-based super-resolution is completely general and could be implemented in other quantum systems where coupling between the modes exists.

The system is sketched in Fig. 1a. Photons are injected to the input of the array, are allowed to propagate, and are then measured by detectors at the output facet of the array. The propagation and coupling between the waveguides are modeled by the Hamiltonian[27]

$$H = \beta \sum a_n^\dagger a_n + C \sum \left( a_{n-1}^\dagger a_n + a_{n+1}^\dagger a_n \right). \quad (1)$$

Here, $a_n^\dagger, a_n$ are the creation and annihilation operators in waveguide $n$, respectively, $\beta$ is the propagation constant (identical to all waveguides) and $C$ is the coupling constant between adjacent waveguides. This Hamiltonian leads to the following Heisenberg equation of motion, which describes the propagation along the $z$ axis

$$i\frac{\partial}{\partial z}a_n^\dagger = -\beta a_n^\dagger - C\left(a_{n-1}^\dagger + a_{n+1}^\dagger\right). \quad (2)$$

The simplest case of propagation in the array occurs when the input is strictly into a single waveguide, say, for example, $n=0$. For this input, the classical solution has a closed-form[31], which coincides with the quantum case when a single photon is injected into the $n=0$ waveguide[5]. The expectation value of observing that photon after propagating a distance z (which can be thought of as the "impulse response" of the quantum system for a single photon input), is shown in Fig. 1a.

Throughout this article, we are interested in recovering an initial quantum state with a fixed number of photons $N$. For simplicity, consider the case where the input state consists of three photons. The quantum state is described by the density matrix of the system. We assume that the basis in which the state is diagonal is known, and this basis will serve as the "sparsity basis" (although in general the sparsity basis can be extracted (learned) from the measurements under certain conditions, or from data with similar features that is often available from other sources[32]). As a first example, consider a state which is diagonal in the Fock basis (another example of an entangled state is given later on). The density matrix takes the form

$$\rho = \sum_i p_i \left|\{n\}^i\right\rangle\left\langle\{n\}^i\right|. \quad (3)$$

Here, $|\{n^i\}\rangle$ is a Fock state with configuration $\{n^i\} = n_1^i n_2^i \cdots n_{N_w}^i$, where the lower index indicates a waveguide number and the upper index $i$ refers to the i-th configuration (so-called configuration index), with $p_i$ describing the probability of that configuration to occur. Since the coefficients $p_i$ are probabilities, they obey $0 \leq p_i \leq 1$, $\sum p_i = 1$, in agreement with the general characterization of quantum states. Accordingly, the density operator satisfies $\rho \geq 0$. An example of such a state is

$$\rho = p|1_2 1_5 1_7\rangle\langle 1_2 1_5 1_7| + (1-p)|2_3 1_{16}\rangle\langle 2_3 1_{16}|. \qquad (4)$$

This density matrix describes the convex sum of two configurations. The first configuration is of probability $p$, and it consists of one photon in waveguide 2, one in waveguide 5 and one in waveguide 7, whereas the second configuration is of probability $1-p$, and has two photons in waveguide 3, and one photon in waveguide 16.

The problem at hand it to recover the initial state at $z = 0$, namely the coefficients $p_i$ from measurements carried out at the output facet after propagating a distance z in the array. Generally, characterization of three-photon state requires three-fold coincidence measurements. Such measurements are described by $\Gamma^{(3)}_{q,r,k} = \text{Tr}\{\rho a_q^\dagger a_r^\dagger a_k^\dagger a_k a_r a_q\}$, where $q, r, k$ are the waveguide indices. Instead, in what follows, we will use only two-fold coincidence measurements, $\Gamma^{(2)}_{q,r} = \text{Tr}\{\rho a_q^\dagger a_r^\dagger a_r a_q\}$, that have the advantages described earlier but are missing considerable information. Substituting Eq. (3) into the expression for two fold coincidences, $\Gamma^{(2)}_{q,r}$, we obtain the relation between the probabilities $p_i$ and the measurements

$$\Gamma_{q,r} = \sum p_i \langle \{n^i\} | a_q^\dagger a_r^\dagger a_r a_q | \{n^i\} \rangle. \quad (5)$$

If we gather the measurements in all the waveguide pairs, casting the problem in a matrix form, we obtain

$$\vec{\Gamma} = M\vec{p}. \quad (6)$$

In this formulation, $\vec{\Gamma} \in \mathbb{R}^{N_m}$ holds all the measurements, $N_m$ is the number of waveguide pairs, $\vec{p} \in \mathbb{R}^{N_b}$ is the sought coefficients (probabilities) vector, $N_b$ is the number of basis vectors, and $M \in \mathbb{R}^{N_m \times N_b}$ is the "sensing matrix" representing the propagation in the array and the relation between the input state and the measurements.

The number of coefficients $N_b = \binom{N_w + 2}{3}$, which is the total number of possible configurations, is derived from the number of photons (3, in this case), and the number of waveguides in the system $N_w$. Unlike the number of basis vectors, which grows with the number of photons, the number of two-fold coincidences $N_m = \binom{N_w + 1}{2}$ depends only on the number of waveguides. Upon examination of the dimensions of the objects in Eq. 6, we learn that $N_m < N_b$ always, hence the problem is inherently non-invertible. For example, if we consider 3 photons in an array of 20 waveguides, we obtain $N_m = 210$, whereas $N_b = 1540$. This is a manifestation of using only two-fold coincidences (instead of the three-fold coincidence which would have made the problem invertible).

To summarize this section, the problem at hand is to find the vector $\vec{p}$ in Eq. 6, which consists of $N_b$ terms from the measurement vector $\vec{\Gamma}$ consisting of $N_m < N_b$ (real) terms, given the matrix $M$. To solve this ill-posed mathematical problem, we need some prior knowledge, which ideally should be rather general. The concept we propose is based on sparsity: the prior knowledge that the initial state has a small number of non-zero elements $p_i$, which physically means that the state is close to a pure state. Such states are common in many experimental scenarios. For instance, if we start with a pure state of a large system and lose access to a small number of degrees of freedom, the resulting reduced density matrix describing the part of the system accessible to measurements is low rank, and this leads to a sparse vector $\vec{p}$ (see Supplementary Information). In a similar vein, a pure state subject to local noise often results in a low rank density matrix[33]. Furthermore, whenever a pure state is subject to a low level of "depolarizing noise" (defined as $\rho \mapsto (1-\lambda)\rho + \frac{\lambda}{N_b} I$ for some $\lambda \in [0,1]$), it is described by a compressible density matrix, which is a density matrix with one (or several, in the case of almost pure states) significant eigenvalues. Such compressible states fall under the scope of our method as well. Finally, bipartite states with low enough rank are also appealing theoretically, since they hold a usable entanglement resource, the so-called distillable entanglement[34].

The usage of sparsity is now being intensively explored in the field of signal processing, typically under the title of Compressed Sensing (CS)[18,35]. For classical signals, CS is a field in information science aimed at reducing the number of measurements required for recovering a signal, given that it is sparse in some basis[36]. An essential condition for CS recovery to work well is that each measurement has to carry information, i.e. an impulse input signal should get 'smeared' as much as possible in the measurement domain. More recently, CS has been brought into the quantum domain for the purpose of reducing the number of measurements necessary in QST[33] and in quantum process tomography[37], enabling much more efficient tomography. The idea of using sparsity has opened the door for a wide range of applications in various fields, from sub-Nyquist sampling[38], to sub-wavelength imaging[39,40] and phase retrieval[40,41]. To distinguish from these, in this work we use sparsity of the sought quantum state in order to recover a three photon state from two-fold correlations, thus achieving quantum super-resolution.

Returning to the problem at hand, we would like to invert Eq.6: find the vector of probabilities $\vec{p}$ given the measurement vector $\vec{\Gamma}$ (which often also contains noise) and the matrix describing the propagation in the waveguide array $M$. In order to overcome the singularity of the problem, we assume that the state is sparse in a known basis, which translates to having a small number of non-zero coefficients $p_i$. It is important to stress that we do not need to know their locations or even their number. The only requirement is that there are few in comparison to the total length of the vector. The recovery of the coefficients is performed algorithmically, based on the coupling between the waveguides and propagation in the array. As known from the field of CS, sparsity-based signal recovery works best if the measurements are carried out in a basis that is least correlated with the basis in which the signal is sparse. It is therefore important to notice that, for a sufficiently long propagation (large value of $Cz$), the input signal is smeared by the impulse response of the system (Fig. 1b). This means that performing measurements at the output of a sufficiently long waveguide array facilitates the use of sparsity-based methods.

Our algorithm is based on Orthogonal Matching Pursuit[42], which is commonly used in sparsity-based approaches, with some modifications derived from our constraints. We note that the standard technique of $l_1$ minimization is not suitable for this problem since every feasible solution has $l_1$ norm of unity, as $\vec{p}$ is a probability vector. Other common methods which utilize sparsity in various ways are applicable here, such as weighted $l_1$ (see further detail in the Supplementary Information).

Examples of sparsity-based reconstructions in the basis of Fock states are presented in Figs. 2a,b. The (simulated) measured data in these examples includes 35dB noise distributed evenly in the measurements. In addition, we assumed that the original state (which is what we wish to recover, the "sought information") includes 2% depolarizing noise, which simulates many physical cases when the preparation of quantum states is imperfect. Figure 2a shows the original signal with the bias resulting from the depolarization noise, which makes the signal compressible (see Supplementary Information). In the recovery process, we wish to obtain the clean signal (without the bias), which is sparse. In Fig. 2b, the original signal is shown without the bias. The figures show the original elements of $\vec{p}$ in bars, and the coefficients recovered by our sparsity-based method from two-fold correlations in squares. The number of elements (sparsity) in the original clean signal is 7, as in the recovered one. Thus, our method deals with the compressible signal and recovers a clean (and sparse) one. These examples highlight the fact that our technique enables virtually perfect recovery of 3-photon states from two-fold coincidence measurements, in the presence of measurement noise and also even when the original quantum state is imperfect. In other words, our sparsity-based method displays robust recovery.

Figure 2c shows the recovery probability for different sparsity levels in a noiseless scenario. The recovery probability, $p = \frac{\#(f > 0.95)}{N_r}$, is defined as the number of recoveries with fidelity higher than 0.95, out of $N_r$ random realizations of the original quantum state (the signal we wish to recover). As expected, the recovery probability decreases as the number of nonzero elements in the signal increases, which means that it is less sparse. Figure 2d shows the performance of our method in terms of fidelity of the two signals $f = \sum \sqrt{p_i \hat{p}_i}$, $p_i$ and $\hat{p}_i$ the elements of the original and recovered signals, respectively, in the presence of various noise levels. The same figure also shows the dependence of the fidelity on sparsity. The method works better when the signal is more sparse, but it yields high fidelity recovery (better than 90%) for up to 20 nonzero terms (out of 1540 possible configurations), under 40 dB noise.

We have thus far demonstrated sparsity-based recovery of 3-photon states from 2-fold coincidence measurements for basis functions that are Fock states. However, the field of quantum information relies heavily on entangled states. It is therefore essential to examine our sparsity-based reconstruction method when the basis includes entangled states. Figure 3 presents exactly that: Figs. 3a,b show examples where the basis consists of spatially-entangled states. As an example, we divide the waveguides between two parties such that "Alice" gets waveguide 7 and the rest of the waveguides belong to "Bob". The basis now consists of the entangled vectors $|\psi\rangle = \frac{|2_3 1_7\rangle + |1_3 2_7\rangle}{\sqrt{2}}, |\psi^\perp\rangle = \frac{|2_3 1_7\rangle + |1_3 2_7\rangle}{\sqrt{2}}$, while the rest of the basis terms are Fock states of all the waveguides other than the pair 3,7. A sparse state in this basis is of the form

$$\rho = p_1 |\psi\rangle\langle\psi| + p_2 |\psi^\perp\rangle\langle\psi^\perp| + \sum_{i=3} p_i |\{n\}^i\rangle\langle\{n\}^i|$$ with a small number of nonzero coefficients $p_i$. This basis has 1540 configurations, which is also the number of possible terms in the "sought signal". In the examples presented in Figs. 3a,b the measurement noise, depolarization noise, and sparsity are the same as in the examples in Fig. 2. The performance of our method, for this basis that includes entangled states, is presented in terms of recovery probability in a noiseless scenario (Fig. 3c) and the average fidelity in various noise levels and sparsity values (Fig. 3d). Clearly, our sparsity-based technique performs as well as it does for the Fock state basis. However, the entanglement adds an unexpected feature: in some of these bases, it yields a two-fold degeneracy which makes it impossible to distinguish between the coefficients of the two specific basis vectors that represent entangled states, $|\psi\rangle, |\psi^\perp\rangle$. This feature could provide a useful avenue for detecting eavesdropping on the information transferred in this system. For example, one could launch such a known state, strictly for the purpose of detecting eavesdropping, and use our sparsity-based reconstruction. The reconstruction would lead to inability to distinguish between two states, whereas eavesdropping would break the degeneracy (while ordinary recovery by using 3-fold coincidence measurements simply recovers the known states without flagging eavesdropping). Alternatively, this degeneracy can be avoided altogether by using a system which varies in $z$ (see further details in the Supplementary Information).

It is important to emphasize that using the sparsity-based methodology presented here is conceptual, not specific to a particular algorithm. Our method is based on using a very general (generic) prior knowledge, namely, that a state is sparse or compressible, in order to solve a noninvertible problem, which is recovering a three-photon state from two-fold correlations. Naturally, other algorithms utilizing sparsity could be used to solve the problem, possibly performing even better than ours (see discussion in the Supplementary Information).

In conclusion, we showed that prior knowledge in the form of sparsity can be used in order to recover a three-photon state from two-fold correlation measurements. This is achieved by coupling the spatial modes through an array of waveguides, in the spirit of CS, and using sparsity-based algorithmics. This idea is readily extendable to recover $N$-photon states from $N-1$ coincidence measurements, because the mathematics is similar (see Supplementary Information for details). Can this idea be extended to cases where the measurements are even more incomplete, for example, recover a 4-photon state from 2-fold coincidence? We leave that for future work although our preliminary results are encouraging. Moreover, we propose extending the sparsity-based ideas to other, closely related, scenarios. For example, in many experiments, number-resolving photon detectors are needed in order to characterize a state. Such detectors are less available and allow lower detection efficiencies. If we replace the number-resolving detectors with ordinary, simple "bucket" detectors, the problem becomes noninvertible. Our preliminary results on this problem indicate that sparsity (i.e., having some structure in the sought state) can be used in order to overcome this problem and allow usage of simple detectors, rendering the usage of number-resolving detectors altogether unnecessary, at least for quantum experiments specifically designed for recovering quantum states.

This work was generously supported by the ICORE program "Circle of Light" administered by the Israel Science Foundation, and by an Advanced Grant from the European Research Council.

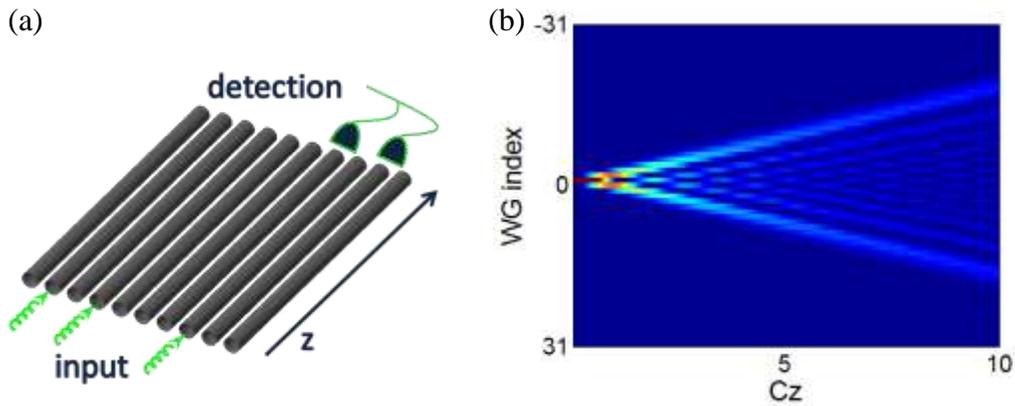

**Figure 1: The physical system and its "quantum impulse response".**

(a) The physical system: an array of evanescently coupled optical waveguides. The quantum state is launched at $z=0$. The photons propagate in the array a distance much larger than the coupling length, such that the information is sufficiently spread among the waveguides in the measurement plane. (b) The probability distribution to measure a photon $\langle \hat{n}_k(z) \rangle$, $C$ the coupling constant between adjacent waveguides, when a single photon is injected into the middle waveguide. It serves as the "impulse response" of the system when the number of waveguides approaches infinity.

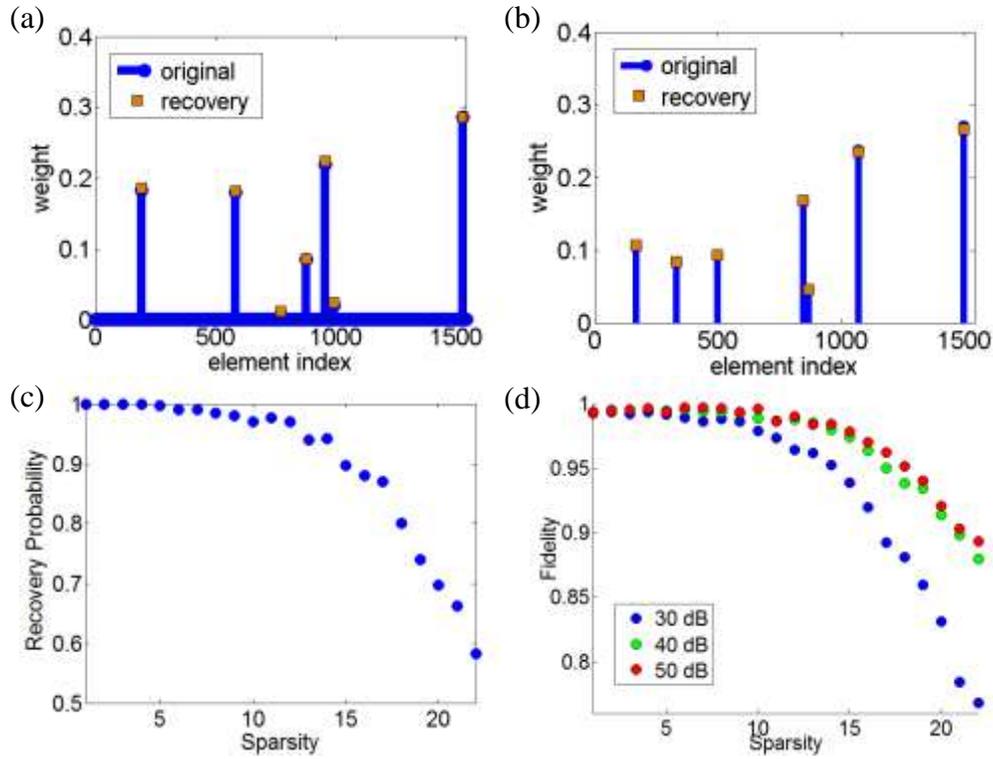

**Figure 2: Sparsity-based recovery on Fock states: examples and performance analysis**

(a) Example for recovery of a state in Fock basis. The original signal has sparsity of 7 (7 nonzero elements). The state is subject to depolarization noise of 2%, which makes the state compressible. The depolarization noise appears in (a) as the thick horizontal line marking an undesired bias. The SNR of the measurements is 35 dB. We wish to recover the true original signal (without the bias noise), which is sparse. As seen in the example, the recovered signal is indeed practically identical to the true original signal. (b) Another example under the same conditions with the depolarization noise not shown. (c) Recovery probability as a function of the number of nonzero elements in the signal. (d) Dependence of the fidelity of recovery on measurement noise.

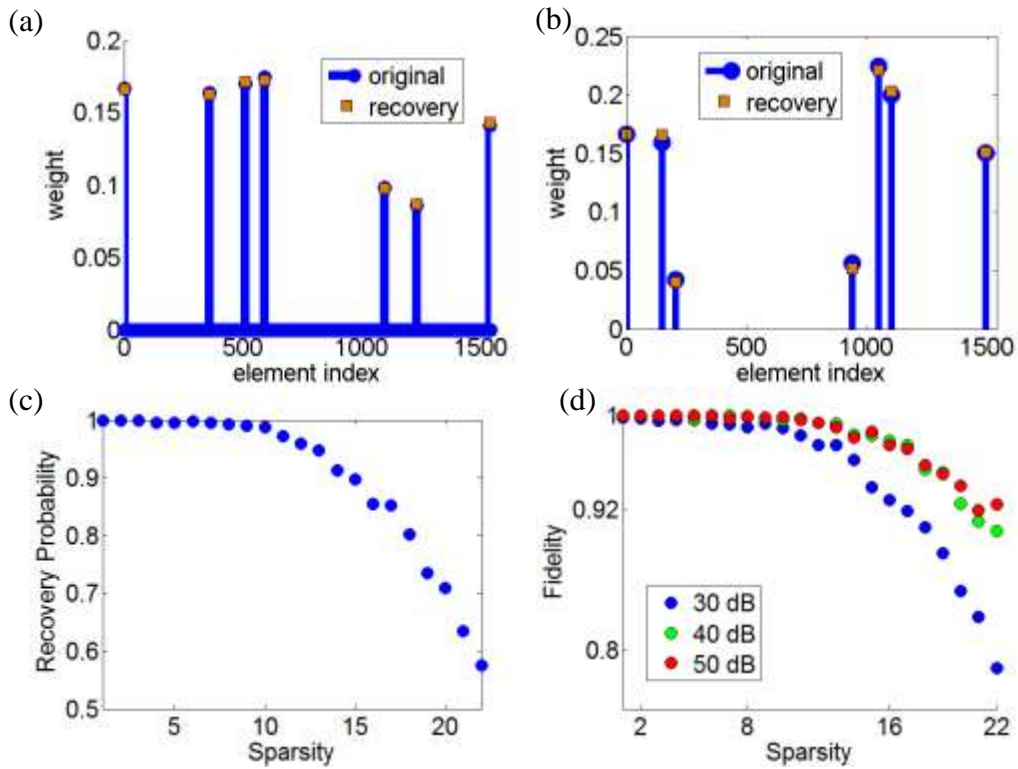

**Figure 3: Sparsity-based recovery on entangled states: examples and performance analysis**

(a) Example for recovery of a state in an entangled basis. The original signal has sparsity of 7 (7 nonzero elements). The state is subject to depolarization noise of 2%, which makes the state compressible. The depolarization noise appears in (a) as the thick horizontal line marking an undesired bias. The SNR of the measurements is 35 dB. We wish to recover the true original signal (without the bias noise), which is sparse. As seen in the example, the recovered signal is indeed practically identical to the true original signal. (b) Another example under the same conditions with the depolarization noise not shown. (c) Recovery probability as a function of the number of nonzero elements in the signal. (d) Dependence of the fidelity of recovery on measurement noise.